\documentclass[%
reprint,
superscriptaddress,
 amsmath,amssymb,
 aps,
]{revtex4-2}

\usepackage{graphicx}% Include figure files
\usepackage{dcolumn}% Align table columns on decimal point
\usepackage{bm}% bold math
\usepackage{color}% color
\usepackage{comment}

\begin{document}

\preprint{APS/123-QED}

\title{Hydrodynamic lift of a two-dimensional liquid domain with odd viscosity}

\author{Yuto Hosaka}\email{yutohosaka24653@gmail.com}
\affiliation{Department of Chemistry, Graduate School of Science, Tokyo Metropolitan University, Tokyo 192-0397, Japan}

\author{Shigeyuki Komura}\email{komura@wiucas.ac.cn}
\affiliation{Department of Chemistry, Graduate School of Science, Tokyo Metropolitan University, Tokyo 192-0397, Japan}
\affiliation{Wenzhou Institute, University of Chinese Academy of Sciences, Wenzhou, Zhejiang 325001, China}
\affiliation{Oujiang Laboratory, Wenzhou, Zhejiang 325000, China}

\author{David Andelman}\email{andelman@tauex.tau.ac.il}
\affiliation{Raymond and Beverly Sackler School of Physics and Astronomy, Tel Aviv University, Ramat Aviv, Tel Aviv 69978, Israel}

\begin{abstract}
We discuss hydrodynamic forces acting on a two-dimensional liquid domain that moves laterally within a supported fluid membrane in the presence of odd viscosity.
Since active rotating proteins can accumulate inside the domain, we focus on the difference in odd viscosity between the inside and outside of the domain.
Taking into account the momentum leakage from a two-dimensional incompressible fluid to the underlying substrate,
we analytically obtain the fluid flow induced by the lateral domain motion, and calculate the drag and lift forces acting on the moving liquid domain.
In contrast to the passive case without odd viscosity, the lateral lift arises in the
active case only when the in/out odd viscosities are different.
The in/out contrast in the odd viscosity leads to nonreciprocal hydrodynamic responses of an active liquid domain.
\end{abstract}

\maketitle

%\tableofcontents

%%%%%%%%%%%%%
\section{Introduction}
%%%%%%%%%%%%%

Biological membranes play an important role in various life-sustaining processes such as the transportation of materials
or the reaction between chemical species, which are essential for cellular metabolism and homeostasis~\cite{albertsbook}.
Biomembranes are composed of two layers of lipid molecules, cholesterol, and various types of proteins that can move laterally
due to the membrane fluidity~\cite{singer1972}.
Since lipid bilayers are extremely thin, as compared to their lateral size, they have been modeled as two-dimensional (2D)
fluids, and their transport properties have been investigated both theoretically and experimentally.
For instance, the drag coefficient of a disk-like domain (protein) moving in a 2D fluid sheet has been studied for various
membrane geometries~\cite{saffman1975,saffman1976,evans1988,ramachandran2010}.
Using fluorescence correlation spectroscopy, Ramadurai \textit{et al.}\ measured the lateral mobility of proteins in lipid bilayers and confirmed a logarithmic dependence of the mobility on the protein size in agreement with predictions~\cite{ramadurai2009}.

In an actual biological environment, the presence of active protein molecules plays an important role
because they induce nonequilibrium hydrodynamic effects to the surrounding
fluid~\cite{mikhailov2015,kapral2016,koyano2016,hosaka2017}.
For example, there are active rotating proteins such as ion pumps, that allow materials to pass
through the membrane~\cite{manneville1999,manneville2001}.
Their inherent nonequilibrium nature due to continuous energy consumption violates the time-reversal symmetry and drives the membrane into out-of-equilibrium situations~\cite{gompper2020}.
In addition, rotating proteins further break the parity symmetry because of their unidirectional motion, so that the membrane with autonomous rotors can be viewed as an active chiral system~\cite{lenz2004,furthauer2012,banerjee2017,oppenheimer2019}.
Moreover, active proteins are often inhomogeneously distributed in the membrane to form active
protein-rich domains that are called lipid rafts~\cite{simons1997,veatch2005,komura2014}.
Due to the presence of such condensed active rotor proteins, biomembranes can be regarded as a heterogeneous active chiral fluid rather than just a uniform and passive 2D fluid.

Active chiral fluids are generally characterized by a peculiar rheological property called \textit{odd viscosity}~\cite{avron1998}, which accounts for the fluid flow perpendicular to the velocity gradient and does not contribute to energy dissipation.
It is known that odd viscosity gives rise to anomalous hydrodynamic phenomena such as surface waves~\cite{abanov2018} or topological edge modes~\cite{souslov2019,tauber2019,tauber2020} at fluid boundaries.
Furthermore, it leads to an instability of a viscous film~\cite{bao2021,mukhopadhyay2021} and asymmetric mobility~\cite{hosaka2021}.
In an incompressible fluid, however, the odd viscosity can be absorbed into the hydrostatic pressure term~\cite{avron1998,banerjee2017} and does not affect the flow profile~\cite{ganeshan2017,souslov2020}.
To clearly see the odd viscosity effect, one should include either the violation of the incompressibility condition or the appropriate boundaries in 2D fluids~\cite{hosaka2021,ganeshan2017}.

To reveal the odd viscosity effect, the hydrodynamic forces acting on various objects have been studied in the presence of odd viscosity~\cite{hosaka2021,lapa2014,ganeshan2017,souslov2020}.
For a laterally moving rigid disk, it was found that odd viscosity causes a hydrodynamic lift force for a compressible 2D fluid~\cite{hosaka2021}.
Moreover, odd viscosity is responsible for the torque acting on objects with time-varying
area such as an expanding bubble with a no-stress boundary condition~\cite{lapa2014,ganeshan2017,souslov2020}.
From the experimental point of view, odd viscosity was measured for a fluid consisting of
self-spinning particles~\cite{soni2019,yang2021}.
Although odd viscosity may exist in biological systems~\cite{banerjee2017,markovich2021},
hydrodynamic responses in heterogeneous active chiral fluids have not been discussed and the role of odd viscosity in biomembranes remains largely unexplored.

In this paper, we discuss the hydrodynamic forces acting on a circular liquid domain that moves laterally in a supported membrane in the presence of odd viscosity~\cite{avron1998}.
To investigate active heterogeneous structures relevant to lipid rafts in biomembranes, we consider a
situation where the odd viscosity is different between the inside and outside of the liquid domain.
Taking into account the momentum leakage from the 2D fluid to the underlying
substrate~\cite{ramachandran2010,seki93,komura95,seki07,sanoop10,sanoop11,sanooppolymer11},
we analytically obtain the velocity field induced by the domain motion and discuss its dependence on the
odd viscosity difference.
We then calculate the drag and lift forces acting on a moving liquid domain.
We show that a dissipationless lift force acting on the domain emerges when only the odd viscosity difference is present, while it vanishes when the odd viscosity is uniform in space.
We further obtain various limiting expressions of the drag and lift coefficients for small and large
domain sizes, which deviate from those obtained for the passive case~\cite{ramachandran2010}.

In the next section, we introduce the hydrodynamic equations for a 2D active chiral fluid with momentum decay
and show a general solution in the presence of odd viscosity.
In Sec.~\ref{sec:velocity_stress}, we obtain the velocity field and stress tensor needed to investigate the
flow profile induced by the domain motion.
In Sec.~\ref{sec:drag_lift}, we calculate the hydrodynamic drag and lift forces acting on the liquid domain and examine their limiting expressions, either by changing the domain size or odd viscosity difference.
Finally, a summary and some further discussions are given in Sec.~\ref{sec:discussion}.

%%%%%%%%%%%%%%%%%%%%%%%%%%%%%%%%%%%
\section{2D Hydrodynamic equations with momentum decay}
%%%%%%%%%%%%%%%%%%%%%%%%%%%%%%%%%%%

\begin{figure}[htb]
\centering
\resizebox{1\columnwidth}{!}{
\includegraphics{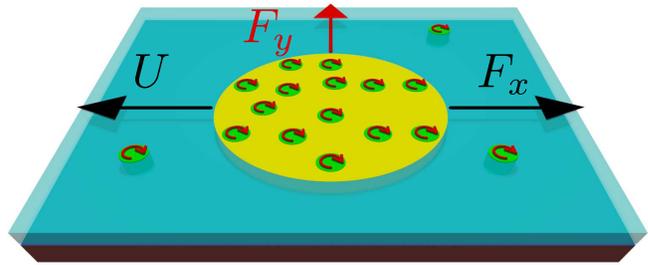}
}
\caption{
Schematic drawing of a fluid membrane (blue), which is flat, thin, incompressible, and supported by a rigid substrate (brown).
The membrane has a 2D even (shear) viscosity $\eta$, odd viscosity $\eta_{\rm o}$, and friction parameter $\lambda$.
A circular liquid domain (yellow) of radius $R$ has a 2D even (shear) viscosity $\eta^\prime$, odd viscosity
$\eta_{\rm o}^\prime$, and friction parameter $\lambda^\prime$.
The odd viscosity reflects the presence of active rotor proteins (green) within the membrane that accumulate inside the liquid domain.
Hence, in general, $\eta_{\rm o}$ can be different from $\eta_{\rm o}^\prime$.
The liquid domain that moves laterally with a velocity $\mathbf{U}=(-U,0)$ experiences a hydrodynamic
force $\mathbf{F}=(F_x,F_y)$, where $F_x$ and $F_y$ are the drag and lateral lift forces, respectively.
}
\label{fig:system}
\end{figure}

Biological membranes are formed as condensed lipid molecules with very small area compressibility~\cite{evans1987} and they have been modeled as incompressible fluids~\cite{evans1988,saffman1975,saffman1976}.
For an incompressible 2D fluid in which momentum is strictly conserved, one cannot obtain
a linear relation between the velocity and viscous force acting on an embedded object.
This is the well-known Stokes' paradox~\cite{lamb1975,landau1987}.
One way to circumvent this problem is to introduce a momentum decay mechanism in
the 2D fluid~\cite{ramachandran2010,evans1988}.
Such a momentum leakage occurs, for example, due to the friction between the supported
membrane and the underlying rigid substrate~\cite{tanaka2005}, as shown in Fig.~\ref{fig:system}.

Let us denote any 2D vector by $\mathbf{r}=(x,y)$ and the 2D velocity by $\mathbf{v}(\mathbf{r})$.
The steady-state linearized hydrodynamic equation for an active chiral fluid in the low Reynolds
number limit can be written as~\cite{ramachandran2010,seki93,komura95,seki07,sanoop10,sanoop11,sanooppolymer11}
\begin{align}
\nabla\cdot\boldsymbol{\sigma} -\lambda\mathbf{v}=0.
\label{eq:balance}
\end{align}
Here, $\nabla=(\partial_x, \partial_y)$ stands for the 2D gradient operator,
$\boldsymbol{\sigma}$ is the 2D fluid stress tensor as given below in Eq.~(\ref{eq:stress}), and $\lambda$ is the friction parameter accounting for the
momentum decay (see also Sec.~\ref{sec:discussion} later for an estimate of $\lambda$).
In addition, we assume that the 2D fluid is incompressible satisfying the condition:
\begin{align}
\nabla\cdot\mathbf{v}=0.
\label{eq:incomp}
\end{align}
The Stokes' paradox can be eliminated in the presence of the momentum decay mechanism, and
one can consistently solve the above hydrodynamic equations under appropriate boundary conditions.

For an incompressible 2D fluid with odd viscosity, the stress tensor is given
by~\cite{banerjee2017,epstein2020,hosaka2021}
\begin{align}
\sigma_{i j} & =-p\delta_{ij} + \eta \left(\partial_{j} v_{i}+\partial_{i} v_{j}\right)
\nonumber \\
& +\frac{1}{2} \eta_{\rm o} \left(\partial_{j} v_{i}^{*}+\partial_{i} v_{j}^{*}+\partial_{j}^{*} v_{i}+\partial_{i}^{*} v_{j}\right),
\label{eq:stress}
\end{align}
where $p$ is the 2D hydrostatic pressure with $\delta_{ij}$ being the Kronecker delta, $\eta$ and $\eta_{\rm o}$ are the 2D even (shear) and odd viscosities, respectively, and $v_i^\ast=\epsilon_{ij}v_j$ and $\partial_i^\ast=\epsilon_{ij}\partial_j$ with $\epsilon_{ij}$ being the 2D Levi-Civita antisymmetric tensor ($\epsilon_{xx}=\epsilon_{yy}=0$ and $\epsilon_{xy}=-\epsilon_{yx}=1$).
Hence, the vector $\mathbf{v}^\ast$ is obtained by rotating $\mathbf{v}$ by $\pi/2$ in a clockwise direction.

In our work, we do not specify the microscopic origin of odd viscosity, but it can be attributed,
for example, to self-spinning objects representing active rotor proteins~\cite{banerjee2017,markovich2021}.
Their continuous energy consumption and autonomous rotation break both time-reversal and parity symmetries, giving rise to odd viscosity in a 2D fluid with active rotor proteins.
Although even viscosity $\eta$ is always positive, odd viscosity $\eta_{\rm o}$ can be either positive
or negative depending on the protein rotational direction.
Substituting Eq.~(\ref{eq:stress}) into Eq.~(\ref{eq:balance}), we obtain the 2D hydrodynamic equation as
\begin{align}
-\nabla p + \eta \nabla^{2} \mathbf{v}+\eta_{\rm o} \nabla^{2} \mathbf{v}^\ast-\lambda \mathbf{v}=0,
\label{eq:stokes}
\end{align}
together with the incompressibility condition of Eq.~(\ref{eq:incomp}).

Within an infinitely extended 2D fluid characterized by $\eta$, $\eta_{\rm o}$, and $\lambda$,
we consider now a circular liquid domain of radius $R$ having a 2D even (shear) viscosity $\eta^\prime$ and friction parameter $\lambda^\prime$~\cite{ramachandran2010}, as schematically presented in Fig.~\ref{fig:system}.
Moreover, we assume that the fluid inside the domain has an odd viscosity $\eta_{\rm o}^\prime$ that
can be different from $\eta_{\rm o}$.
The difference in the odd viscosities, $\eta_{\rm o}\neq\eta_{\rm o}^\prime$, reflects the fact that active rotor proteins can accumulate and have a denser concentration in the liquid domain~\cite{simons1997}.
In general, both $\eta_{\rm o}$ and $\eta_{\rm o}^\prime$ can be either positive or negative.
Notice that the domain perimeter is assumed to be impermeable, so that the fluids inside and outside the domain do not mix with each other~\cite{ramachandran2010}.
In addition, we assume that the deformation of the circular liquid domain can be neglected.
This is justified when the line tension at the domain boundary is large enough compared to the viscous force~\cite{landau1987,ramachandran2010}.

Throughout this work, we adopt the notation convention that quantities with prime refer to those inside the domain, while quantities without prime correspond to those outside the domain.
Any 2D fluid velocity can be expressed as the sum of a gradient of a scalar potential $\phi$ and a curl of a vector potential $\mathbf{A} = (0, 0, A)$, where the $z$-component, $A$, corresponds to the stream function~\cite{lamb1975,ramachandran2010}.
Then, the 2D velocities outside/inside the domain are expressed as
\begin{align}
\mathbf{v}=-\nabla\phi +\nabla\times \mathbf{A},~~~~~
\mathbf{v}^\prime=-\nabla\phi^\prime +\nabla\times \mathbf{A}^\prime.
\label{eq:v}
\end{align}
Substituting Eq.~(\ref{eq:v}) into Eq.~(\ref{eq:incomp}), we obtain
\begin{align}
\nabla^2\phi=0,~~~~~
\nabla^2\phi^\prime=0,
\label{eq:lap}
\end{align}
which are the 2D Laplace equations.

One can also show that Eq.~(\ref{eq:stokes}) is satisfied if the outside/inside pressures are given by
\begin{align}
p = \eta\kappa^2 \phi-\eta_{\rm o}\kappa^2A ,~~~~~
p^\prime = \eta^\prime\kappa^{\prime2} \phi^\prime-\eta_{\rm o}^\prime\kappa^{\prime2}A^\prime,
\label{eq:pressure}
\end{align}
while $A$ and $A^\prime$ obey the 2D Helmholtz equations:
\begin{align}
(\nabla^2-\kappa^2)A=0,~~~~~
(\nabla^2-\kappa^{\prime2})A^\prime=0.
\label{eq:helm}
\end{align}
Here, we have defined the inverse hydrodynamic screening lengths for the outside/inside fluids as $\kappa=(\lambda/\eta)^{1/2}$ and $\kappa^\prime=(\lambda^\prime/\eta^\prime)^{1/2}$.
As seen in Eq.~(\ref{eq:pressure}), the effect of odd viscosity can be taken into account through
the modified pressure~\cite{banerjee2017,ganeshan2017,avron1998,lapa2014}, reflecting the fact that the odd viscosity does not
contribute to the dissipation.
In the next section, we shall derive the solutions to Eqs.~(\ref{eq:lap}) and (\ref{eq:helm}) under the
appropriate boundary conditions for a laterally moving liquid domain.

%%%%%%%%%%%%%%%%%%%%%%%%%%%%%%%%%%%%%
\section{The velocity field of a moving liquid domain}
%%%%%%%%%%%%%%%%%%%%%%%%%%%%%%%%%%%%%
\label{sec:velocity_stress}

%%%%%%%%%%%%%%%%%%%%%%%
\subsection{Velocity and stress tensor}

For convenience, we use the 2D polar coordinates $(r,\theta)$ defined by
$x=r \cos\theta$ and $y=r \sin\theta$ with the origin fixed at the domain center.
First, we consider the region outside the domain $(r>R)$.
Under the condition that the velocity and pressure vanish at large distances $r\to\infty$, we write
down the solutions to Eqs.~(\ref{eq:lap}) and (\ref{eq:helm}) as follows:
\begin{align}
\phi =\frac{C_{1}}{r}\cos\theta+\frac{C_3}{r}\sin \theta,
\label{eq:phi}
\end{align}
\begin{align}
A =C_{2} K_{1}(\kappa r) \sin \theta+C_{4} K_{1}(\kappa r) \cos \theta.
\label{eq:A}
\end{align}
Here, $C_1,\cdots,C_4$ are unknown coefficients that will be determined from the boundary conditions,
and $K_1(z)$ is the first-order modified Bessel function of the second kind~\cite{abramowitzhandbook}.
%Notice that $C_3$ and $C_4$ in Eqs.~(\ref{eq:phi}) and (\ref{eq:A}) are nonzero when the odd viscosity is
%present, whereas they vanish for the passive case~\cite{ramachandran2010}.
%This is because the odd viscosity contributes to the fluid stress perpendicular to the velocity gradient, as can be recognized in Eq.~(\ref{eq:stress}).

From Eq.~(\ref{eq:v}), the radial and tangential components of the velocity for $r>R$ are given by
\begin{align}
v_r&=
\left[
\frac{C_3}{r^2}-\frac{C_4}{r}K_1(\kappa r)\right]\sin\theta
+\left[\frac{C_1}{r^2}+\frac{C_2}{r}K_1(\kappa r)\right]\cos\theta,
\label{eq:vr}
\end{align}
and
\begin{align}
v_\theta &=
\left[\frac{C_1}{r^2}+C_2\kappa K_0( \kappa r)+\frac{C_2}{r}K_1(\kappa r)
\right]\sin\theta\nonumber\\
&+
\left[-\frac{C_3}{r^2}+C_4\kappa K_0( \kappa r)+\frac{C_4}{r}K_1(\kappa r)
\right]\cos\theta,
\label{eq:vt}
\end{align}
respectively.
Then, with the use of Eq.~(\ref{eq:stress}), the two components of the stress tensor can be obtained as
\begin{align}
\sigma_{rr} &=-\left[
\eta\left( \frac{4C_3}{r^3}+\frac{C_3\kappa^2}{r}-\frac{2C_4\kappa}{r}K_2(\kappa r) \right)\right.\nonumber \\
&\left.+\eta_{\rm o}\left( \frac{4C_1}{r^3}+\frac{2C_2\kappa}{r}K_2(\kappa r) \right)\right]\sin\theta\nonumber\\
&-\left[
\eta\left( \frac{4C_1}{r^3}+\frac{C_1\kappa^2}{r}+\frac{2C_2\kappa}{r}K_2(\kappa r) \right)\right.\nonumber \\
&\left.+\eta_{\rm o}\left( -\frac{4C_3}{r^3}+\frac{2C_4\kappa}{r}K_2(\kappa r) \right)\right]\cos\theta,
\label{eq:srr}
\end{align}
and
\begin{align}
 \sigma_{r\theta} &=
- \left[ \eta\left( \frac{4C_1}{r^3}  +  C_2 \kappa ^2 K_1(\kappa r)
+\frac{2C_2\kappa}{r} K_2(\kappa r)
\right)  \right.\nonumber\\
&\left.+\eta_{\rm o}\left(- \frac{4C_3}{r^3} +\frac{2 C_4 \kappa}{r}K_2(\kappa r)
 \right)\right] \sin\theta \nonumber \\
&-\left[
\eta\left( -\frac{4C_3}{r^3}
+\frac{2 C_4 \kappa  }{r}K_2(\kappa r)+C_4  \kappa ^2 K_1(\kappa r)
 \right) \right.\nonumber\\
&\left.+\eta_{\rm o}\left(-\frac{4C_1}{r^3}- \frac{2C_2 \kappa }{r}K_2(\kappa r)
\right) \right]\cos\theta.
\label{eq:srt}
\end{align}

Inside the domain $(r < R)$, on the other hand, the solutions to Eqs.~(\ref{eq:lap}) and (\ref{eq:helm}) under the condition that they are finite at the origin $(r=0)$ are given by
\begin{align}
\phi^{\prime} & =C_{1}^{\prime} r \cos \theta+C_{3}^{\prime} r \sin \theta,
\label{eq:phip}
\end{align}
\begin{align}
A^{\prime}=C_{2}^{\prime} I_{1}\left(\kappa^{\prime} r\right) \sin \theta+C_{4}^{\prime} I_{1}\left(\kappa^{\prime} r\right) \cos \theta.
\label{eq:Ap}
\end{align}
Here, $C_1^\prime,\cdots,C_4^\prime$ are the other unknown coefficients, and $I_1(z)$ is the first-order modified Bessel function of the first kind~\cite{abramowitzhandbook}.
Although the general solutions to Eqs.~(\ref{eq:lap}) and (\ref{eq:helm}) for $\phi,\phi^\prime,A$, and $A^\prime$ can be expressed as a series expansion in terms of $r$ and $\theta$, we have kept only the smallest number of terms satisfying the boundary conditions that will be discussed in the next subsection.

\begin{figure*}[thb]
\centering
\includegraphics[scale=0.675]{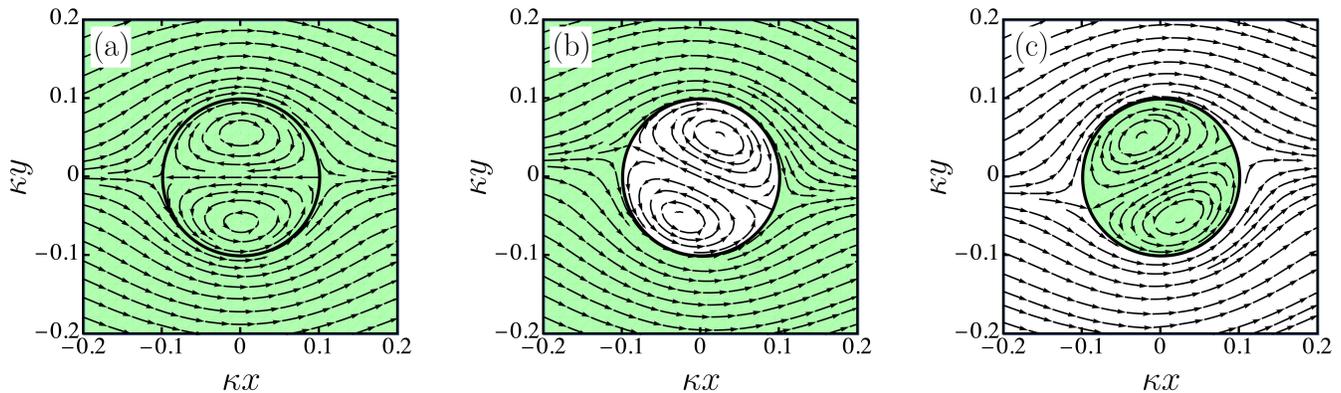}
\caption{
Streamlines (black arrows) of the fluid velocity, $\mathbf{v}-\mathbf{U}$, as a function of $\kappa x$ and
$\kappa y$ when (a) $\eta_{\rm o}=\eta_{\rm o}^\prime=\eta$ (uniform odd viscosity),
(b) $\eta_{\rm o}=\eta$ and $\eta_{\rm o}^\prime=0$ (vanishing odd viscosity inside the domain), and (c) $\eta_{\rm o}=0$ and $\eta_{\rm o}^\prime=\eta$ (vanishing odd viscosity outside the domain) [see Eqs.~(\ref{eq:vr}), (\ref{eq:vt}), (\ref{eq:vrp}), and (\ref{eq:vtp})].
The green (light gray) region represents fluids with nonvanishing odd viscosity, while the white region represents vanishing odd viscosity.
We also have chosen $\eta=\eta^\prime$, $\lambda=\lambda^\prime$, and $\epsilon=\kappa R=0.1$.
The domain moves laterally in the negative $x$-direction with a velocity $\mathbf{U}=(-U,0)$.
The circular black line represents the domain perimeter.
}
\label{fig:v_x_y}
\end{figure*}

Then, the corresponding radial and tangential components of the velocity for $r<R$ become
\begin{align}
v_r^\prime&=
\left[
-C_3^\prime-\frac{C_4^\prime}{r}I_1(\kappa^\prime r)
\right]\sin\theta
-\left[C_1^\prime-\frac{C_2^\prime}{r}I_1(\kappa^\prime r)\right]\cos\theta,
\label{eq:vrp}
\end{align}
and
\begin{align}
v_\theta^\prime &=
\left[
C_1^\prime-C_2^\prime\kappa^\prime I_0(\kappa^\prime r) +\frac{C_2^\prime}{r}I_1(\kappa^\prime r)
\right]\sin\theta\nonumber\\
&+
\left[-C_3^\prime-C_4^\prime\kappa^\prime I_0( \kappa^\prime r)+\frac{C_4^\prime}{r}I_1(\kappa^\prime r)
\right]\cos\theta,
\label{eq:vtp}
\end{align}
respectively, and the two components of the stress tensor are given by
\begin{align}
\sigma_{rr}^\prime &=-\left[
\eta^\prime\left( C_3^\prime\kappa^{\prime2}r+\frac{2C_4^\prime\kappa^\prime}{r}I_2(\kappa^\prime r) \right)\right.\nonumber \\
&\left.-\eta_{\rm o}^\prime\frac{2C_2^\prime\kappa^\prime}{r}I_2(\kappa^\prime r) \right]\sin\theta\nonumber\\
&-\left[
\eta^\prime\left( C_1^\prime\kappa^{\prime2} r -\frac{2C_2^\prime\kappa^\prime}{r}I_2(\kappa^\prime r) \right)\right.\nonumber \\
&\left.-\eta_{\rm o}^\prime \frac{2C_4^\prime\kappa^\prime}{r}I_2(\kappa^\prime r)\right]\cos\theta,
\label{eq:srrp}
\end{align}
and
\begin{align}
\sigma_{r\theta}^\prime &=
- \left[ \eta^\prime C_2^\prime \left( \kappa^{\prime2} I_1(\kappa^\prime r)
-\frac{2\kappa^\prime}{r}I_0(\kappa^\prime r)
+\frac{4  }{r^2}I_1(\kappa^\prime r) \right)  \right.\nonumber\\
&\left.+\eta_{\rm o}^\prime C_4^\prime\left(-\frac{2 \kappa^\prime}{r}I_0(\kappa^\prime r)
+\frac{4}{r^2}I_1(\kappa^\prime r) \right)\right] \sin\theta \nonumber \\
&-\left[
\eta^\prime C_4^\prime\left(
-\frac{2 \kappa^\prime}{r}I_0(\kappa^\prime r)
+\kappa^{\prime2} I_1(\kappa^\prime r)
+\frac{4 }{r^2}I_1( \kappa^\prime r) \right) \right.\nonumber\\
&\left.+\eta_{\rm o}^\prime C_2^\prime\left(
\frac{2 \kappa^\prime}{r}I_0(\kappa^\prime r)
-\frac{4}{r^2}I_1(\kappa^\prime r)\right) \right]\cos\theta.
\label{eq:srtp}
\end{align}
These velocities and stress tensor components for the inside and outside of the domain should be connected through the appropriate boundary conditions at the domain perimeter.

%%%%%%%%%%%%%%%%%%%%%%%%%%%%%%%%%%%%%%
\subsection{Boundary conditions at the liquid domain perimeter}

As mentioned in the previous section, we consider the situation in which the liquid domain is
laterally moving with a constant velocity $\mathbf{U}=(-U,0)$.
At $r=R$, the radial component of the fluid velocity should be equal to the domain velocity, while the
tangential components of the fluid velocity and the stress tensor should be continuous~\cite{landau1987,ramachandran2010}.
These conditions are written as
\begin{align}
v_{r} &=-U\cos\theta,
\label{bcvr} \\
v_{r}^{\prime} &=-U\cos\theta,
\label{bcvrp} \\
v_{\theta} &=v_{\theta}^{\prime},
\label{bcvtheta} \\
\sigma_{r \theta} &=\sigma_{r \theta}^{\prime}.
\label{bcsigmart}
\end{align}

Since we consider the circular liquid domain without deformation, there exists a finite line tension at the domain boundary, which dominates over a viscous force.
The line tension gives rise to the 2D Laplace pressure at the domain perimeter, so that the normal stress condition inside and outside the domain is automatically satisfied~\cite{landau1987,ramachandran2010}.
Hence, one does not need the condition, $\sigma_{rr}=\sigma_{rr}^\prime$, in addition to Eqs.~(\ref{bcvr})-(\ref{bcsigmart}).
Using the above boundary conditions, we can determine the eight coefficients
$C_1, \cdots, C_4, C_1^\prime, \cdots, C_4^\prime$, whose explicit expressions are provided in Appendix~\ref{app:c}.
Since each of Eqs.~(\ref{bcvr})-(\ref{bcsigmart}) includes both $\sin\theta$ and $\cos\theta$ that are orthogonal to each other, one boundary condition provides two constraints.
Therefore, the four boundary conditions lead to eight constraints that are sufficient to determine the eight unknown coefficients.

Notice that for the passive case without odd viscosity $(\eta_{\rm o}=\eta_{\rm o}^\prime=0)$, the coefficients, $C_3, C_4, C_3^\prime,$ and $C_4^\prime$, in Eqs.~(\ref{eq:phi}), (\ref{eq:A}), (\ref{eq:phip}), and (\ref{eq:Ap}) are not required to satisfy the boundary conditions of Eqs.~(\ref{bcvr})-(\ref{bcsigmart})~\cite{ramachandran2010}.
This is because the odd viscosity contributes to the fluid stress perpendicular to the velocity gradient, as can be recognized in Eq.~(\ref{eq:stress}).
More details on the passive case will be summarized in Appendix~\ref{app:sanoop}.

%%%%%%%%%%%%%%%
\subsection{Flow profile}

\begin{figure}[htb]
\centering
\includegraphics[scale=0.45]{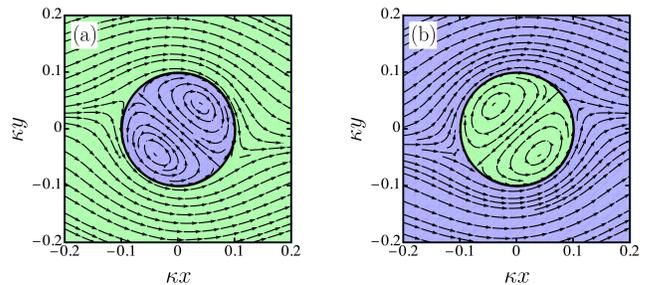}
\caption{
Streamlines (black arrows) of the fluid velocity, $\mathbf{v}-\mathbf{U}$, as a function of $\kappa x$ and
$\kappa y$ when (a) $\eta_{\rm o}=-\eta_{\rm o}^\prime=\eta$ and (b) $-\eta_{\rm o}=\eta_{\rm o}^\prime=\eta$ [see the caption of Fig.~\ref{fig:v_x_y} for the other conditions].
The green (light gray) region represents fluids with positive odd viscosity, while the blue (gray) region represents negative odd viscosity.
}
\label{fig:v_x_y_2}
\end{figure}

Having fixed all the coefficients in Eqs.~(\ref{eq:vr}), (\ref{eq:vt}), (\ref{eq:vrp}), and (\ref{eq:vtp}), we proceed by investigating the fluid flow induced by the lateral translational motion of the liquid domain.
For the sake of simplicity, we assume $\eta=\eta^\prime$ and $\lambda=\lambda^\prime$
(or equivalently $\kappa=\kappa^\prime$).
In Fig.~\ref{fig:v_x_y}, the velocity field $\mathbf{v}-\mathbf{U}$ is plotted for (a) $\eta_{\rm o}=\eta_{\rm o}^\prime=\eta$ (uniform odd viscosity),
(b) $\eta_{\rm o}=\eta$ and $\eta_{\rm o}^\prime=0$
(vanishing odd viscosity inside the domain), and
(c) $\eta_{\rm o}=0$ and $\eta_{\rm o}^\prime=\eta$ (vanishing odd viscosity outside the domain).
In Fig.~\ref{fig:v_x_y_2}, we also plot $\mathbf{v}-\mathbf{U}$ for (a) $\eta_{\rm o}=-\eta_{\rm o}^\prime=\eta$ and (b) $-\eta_{\rm o}=\eta_{\rm o}^\prime=\eta$.
In these plots, the domain size is fixed to $\kappa R=0.1$ (circular black line).

When the odd viscosity is spatially uniform $(\eta_{\rm o}=\eta_{\rm o}^\prime)$, as in Fig.~\ref{fig:v_x_y}(a),
we see that the flow streamlines induced by the domain motion are symmetric with respect to the direction of motion.
Such a symmetric profile is also seen for the passive case in which odd viscosity does not
exist~\cite{ramachandran2010}.
When $\eta_{\rm o}\neq\eta_{\rm o}^\prime$, as in Figs.~\ref{fig:v_x_y}(b) and \ref{fig:v_x_y}(c), the flow inside
the domain is rotated with respect to the $x$-axis and the above symmetry breaks down.
When $\eta_{\rm o}/\eta_{\rm o}^\prime<0$, as in Fig.~\ref{fig:v_x_y_2}, the flow inside the domain is more rotated compared to Figs.~\ref{fig:v_x_y}(b) and \ref{fig:v_x_y}(c).
This implies that the negative odd viscosity enhances the rotation in the flow field.
Figure~\ref{fig:v_x_y}(c) is relevant to a lipid domain enriched with active rotor proteins, while Fig.~\ref{fig:v_x_y_2} represents active proteins rotating oppositely inside and outside the domain.
In the next section, we show that such a flow-field asymmetry leads to a lateral lift force
acting on the domain.

%%%%%%%%%%%%%%%%%%%%%%%%%%%%%%%%%%%%%
\section{Hydrodynamic forces acting on a moving liquid domain}
%%%%%%%%%%%%%%%%%%%%%%%%%%%%%%%%%%%%%
\label{sec:drag_lift}

%%%%%%%%%%%%%%%%%%%
\subsection{Drag and lift forces}

For a liquid domain laterally moving with a velocity $\mathbf{U} = (-U , 0)$, the forces acting in the
$x$- and $y$- directions, $\mathbf{F}=(F_x,F_y)$, are given by~\cite{landau1987,ramachandran2010}
\begin{align}
F_x &= R \int_{0}^{2 \pi}d\theta\, \left(\sigma_{r r} \cos \theta-\sigma_{r \theta} \sin \theta\right)\nonumber \\
&= \pi\eta (\kappa R)^2 \left[-\frac{C_1}{R^2}+\frac{C_2K_1(\kappa R)}{R}\right],
\label{eq:Fx}
\end{align}
and
\begin{align}
F_y &= R \int_{0}^{2 \pi} d\theta\, \left(\sigma_{r r} \sin \theta+\sigma_{r \theta} \cos \theta\right)\nonumber\\
&= -\pi\eta(\kappa R)^2 \left[ \frac{C_3}{R^2}+ \frac{C_4K_1(\kappa R)}{R}\right],
\label{eq:Fy}
\end{align}
respectively.
In the above, the already determined coefficients $C_1, \cdots, C_4$ are substituted as given in Appendix~\ref{app:c}.
In addition, the full expressions of $F_x$ and $F_y$ are also given in Appendix~\ref{app:c}.

\begin{figure}[htb]
\centering
\includegraphics[scale=0.55]{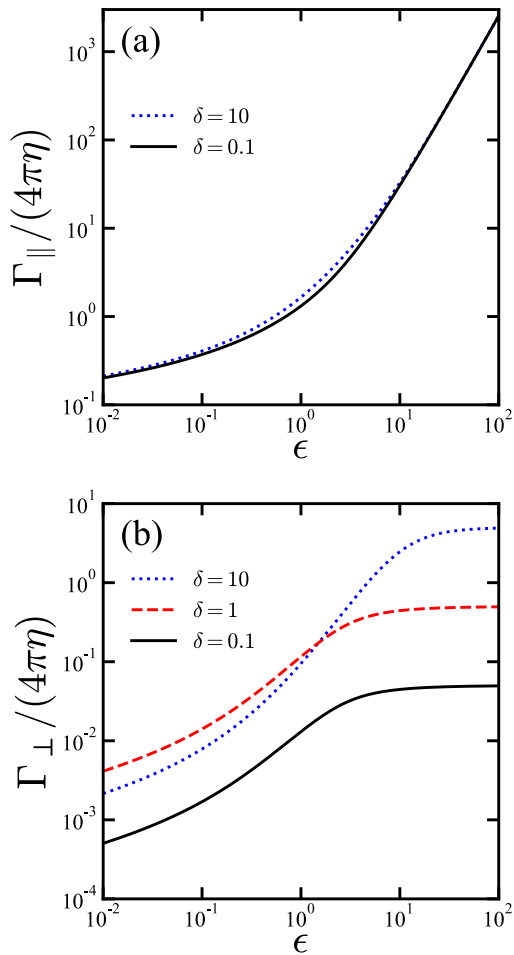}
\caption{
Plots of (a) the rescaled drag coefficient $\Gamma_\|$ and (b) the rescaled lift coefficient $\Gamma_\perp$
as a function of the rescaled domain radius $\epsilon=\kappa R$ for various values of the odd viscosity difference
$\delta=(\eta_{\rm o}-\eta_{\rm o}^\prime)/\eta$.
In (a), $\delta=0.1$ and $10$ are presented by the solid black and dotted blue lines, respectively.
In (b), $\delta=0.1,1$, and $10$ are presented by the solid black, dashed red, and dotted blue lines, respectively.
}
\label{fig:drag_lift}
\end{figure}

For the sake of simplicity, we consider as before the case $\eta=\eta^\prime$ and
$\lambda=\lambda^\prime$ (or equivalently $\kappa=\kappa^\prime$) in Eqs.~(\ref{eq:Fx}) and (\ref{eq:Fy}).
We introduce a dimensionless domain radius, $\epsilon\equiv\kappa R$, and the arguments of the modified Bessel functions are omitted as in $K_n = K_n(\epsilon)$ and
$I_n = I_n(\epsilon)$ to keep the notations more compact.
Then, the expressions for the drag coefficient $\Gamma_\|=F_x/U$ and the lateral lift coefficient
$\Gamma_\perp=F_y/U$ become
\begin{align}
\frac{\Gamma_\|}{4\pi\eta}&=
\frac{\epsilon^2}{4}
+\frac{\epsilon K_1}{K_0}
\left[1- \frac{\epsilon^2(K_0I_1+K_1I_2)K_1I_2}
{\epsilon^2(K_0I_1+K_1I_2)^2+4(\delta K_0I_2)^2}
\right],
\label{eq:drag}
\end{align}
and
\begin{align}
\frac{\Gamma_\perp}{4\pi\eta}&=
\frac{2\delta(\epsilon K_1I_2)^2}
{\epsilon^2(K_0I_1+K_1I_2)^2+4(\delta K_0I_2)^2}.
\label{eq:lift}
\end{align}
In the above, we have introduced the dimensionless difference in odd viscosity, $\delta$, between the inside and outside of the domain
\begin{align}
\delta=\frac{\eta_{\rm o}-\eta_{\rm o}^\prime}{\eta}.
\end{align}
Equations~(\ref{eq:drag}) and (\ref{eq:lift}) are the main results of our work.

Both the drag $\Gamma_\|$ and lift $\Gamma_\perp$ coefficients depend on the odd viscosity difference $\delta$, and are even and odd functions of $\delta$, respectively.
As the domain moves in the negative $x$-direction, it exhibits a lateral lift motion along the $y>0$ direction when $\delta>0$, and also along the $y<0$ direction for $\delta<0$.
Notice that the passive case without odd viscosity $(\eta_{\rm o}=\eta_{\rm o}^\prime=0)$
is recovered by setting $\delta=0$~\cite{ramachandran2010} [see Eq.~(\ref{eq:drag0}) in Appendix~\ref{app:sanoop} for the specific expression].
For the uniform case with $\eta_{\rm o}=\eta_{\rm o}^\prime \neq 0$ or $\delta =0$, the drag coefficient $\Gamma_\|$ reduces to that of the passive case~\cite{ramachandran2010},
whereas the lift coefficient $\Gamma_\perp$ vanishes.
Since the lift force does not exist for the passive
case~\cite{ramachandran2010,evans1988,saffman1975,saffman1976}, the finite lift force reflects not only the existence of odd viscosity, but also its difference $(\delta\neq0)$ between
the inside and outside of the domain.

%%%%%%%%%%%%%%%%%%%%%%%%%%%%%%%%
\subsection{Dependence on the domain size $\epsilon$}

To discuss the dependence of the drag coefficient $\Gamma_\|$ and the lateral
lift coefficient $\Gamma_\perp$ on the domain size $\epsilon$ for arbitrary $\delta$, it is useful to obtain their asymptotic
expressions in the small and large $\epsilon$ limits.
The dependence on $\delta$ will be separately discussed in the next subsection.
For $\epsilon\ll1$, they become
\begin{align}
\frac{\Gamma_\|}{4\pi\eta}&\approx
\frac{4[\ln(2/\epsilon)-\gamma+1/4]+\delta^2[\ln(2/\epsilon)-\gamma]}
{4[\ln(2/\epsilon)-\gamma+1/4]^2+\delta^2[\ln(2/\epsilon)-\gamma]^2},
\label{eq:drags}
\end{align}
and
\begin{align}
\frac{\Gamma_\perp}{4\pi\eta}&\approx
\frac{\delta}
{8[\ln(2/\epsilon)-\gamma+1/4]^2+2\delta^2[\ln(2/\epsilon)-\gamma]^2},
\label{eq:lifts}
\end{align}
where $\gamma \approx 0.5772$ is Euler's constant.
Hence, both $\Gamma_\|$ and $\Gamma_\perp$ depend only logarithmically on the rescaled domain
size $\epsilon$.
In the opposite limit of $\epsilon\gg1$, the asymptotic expressions become
\begin{align}
\frac{\Gamma_\|}{4\pi\eta}&\approx\frac{\epsilon^2}{4},
\label{eq:dragl}
\end{align}
and
\begin{align}
\frac{\Gamma_\perp}{4\pi\eta}&\approx \frac{\delta}{2}.
\label{eq:liftl}
\end{align}
Here, $\Gamma_\|$ is proportional to $\epsilon^{2}$ and independent of $\delta$, while $\Gamma_\perp$ is independent of
$\epsilon$ and is determined solely by $\delta$.

In Fig.~\ref{fig:drag_lift}, we plot $\Gamma_\|$ and $\Gamma_\perp$ of Eqs.~(\ref{eq:drag}) and (\ref{eq:lift}),
respectively, as a function of the rescaled domain size $\epsilon=\kappa R$ for various values of $\delta$.
These plots are consistent with the above asymptotic behaviors of $\Gamma_\|$ and $\Gamma_\perp$.
We also see that the crossover between the two limiting cases is reasonably given for $\epsilon\approx1$.
In Fig.~\ref{fig:drag_lift}(a), $\Gamma_\|$ is slightly larger when $\delta$ is increased, whereas it hardly depends on
$\delta$ for larger $\epsilon$.
In Fig.~\ref{fig:drag_lift}(b), we see that the lift coefficient $\Gamma_\perp$ increases logarithmically for
$\epsilon \ll 1$, while it becomes independent of the domain size for $\epsilon\gg1$.

\begin{figure}[htb]
\centering
\includegraphics[scale=0.55]{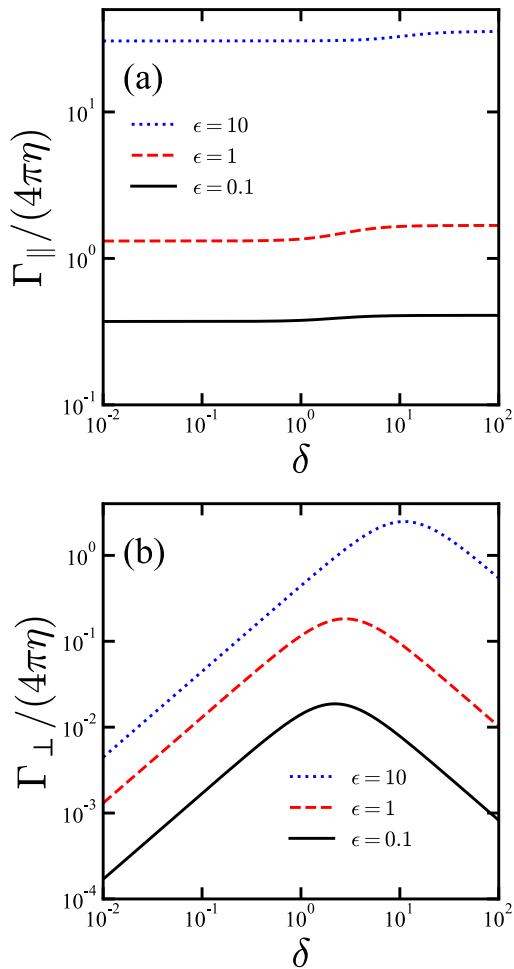}
\caption{
Plots of (a) the rescaled drag coefficient $\Gamma_\|$ and (b) the rescaled lift coefficient $\Gamma_\perp$
as a function of the odd viscosity difference $\delta=(\eta_{\rm o}-\eta_{\rm o}^\prime)/\eta$ for various
values of the rescaled domain radius $\epsilon=\kappa R$.
In both plots, $\epsilon=0.1,1$, and $10$ are presented by the solid black, dashed red, and dotted blue lines,
respectively.
}
\label{fig:lift_delta}
\end{figure}

Let us discuss the physical interpretation of the above limiting behaviors of
$\Gamma_\|$ and $\Gamma_\perp$~\cite{diamant2009,sanooppolymer11}.
The momentum in the 2D fluid is conserved over distances smaller than the hydrodynamic screening length,
$r \ll \kappa^{-1}$, and the stress decays as $1/r$ due to the momentum conservation.
Since the stress scales as $\sigma \sim \eta v/r$, we have
$v \sim 1/\eta$~\cite{diamant2009}.
This explains the weak (logarithmic) size dependence of $\Gamma_\|$ and $\Gamma_\perp$ in
Eqs.~(\ref{eq:drags}) and (\ref{eq:lifts}), respectively.
For larger length scales, $r \gg \kappa^{-1}$, the momentum is not conserved, and the only
contribution to the velocity is from mass conservation.
In a 2D fluid, a mass monopole (source) will create a velocity that decays as $1/r$~\cite{diamant2009}.
Hence, the velocity due to a mass dipole (source and sink) decays as $1/r^2$, explaining the scaling
$\Gamma_\| \sim \epsilon^2$ in Eq.~(\ref{eq:dragl}).
Such a strong size dependence is not observed for $\Gamma_\perp$ in Eq.~(\ref{eq:liftl}) as the friction parameter $\lambda$ does not cause any momentum leakage along the rotated velocity $\mathbf{v}^\ast$.

%%%%%%%%%%%%%%%%%%%%%%%%%%%%%%%%%%%%%%
\subsection{Dependence on the odd viscosity difference $\delta$}

Next, we show how $\Gamma_\|$ and $\Gamma_\perp$ depend on the odd viscosity difference $\delta$
for arbitrary $\epsilon$.
The asymptotic expressions of Eqs.~(\ref{eq:drag}) and (\ref{eq:lift}) for $|\delta|\ll1$ are
\begin{align}
\frac{\Gamma_\|}{4\pi\eta}&\approx
\frac{\epsilon^2}{4}+\frac{\epsilon K_1I_1}{K_0I_1+K_1I_2},
\label{dragsmalldelta}
\end{align}
and
\begin{align}
\frac{\Gamma_\perp}{4\pi\eta}&\approx
2\delta \left( \frac{ K_1I_2}{K_0I_1+K_1I_2}\right)^2,
\label{liftsmalldekta}
\end{align}
showing that $\Gamma_\|$ is independent of $\delta$ and
$\Gamma_\perp$ is proportional to only $\delta$.
As mentioned before, Eq.~(\ref{dragsmalldelta}) coincides with the passive drag coefficient of a liquid
domain~\cite{ramachandran2010} [see Eq.~(\ref{eq:drag0})].

When $|\delta|\gg1$, on the other hand, we obtain
\begin{align}
\frac{\Gamma_\|}{4\pi\eta}&\approx
\frac{\epsilon^2}{4}+
\frac{\epsilon K_1}{K_0},
\label{draglargedelta}
\end{align}
and
\begin{align}
\frac{\Gamma_\perp}{4\pi\eta}&\approx
\frac{1}{2\delta} \left(\frac{\epsilon K_1}{K_0}\right)^2.
\label{liftlargedelta}
\end{align}
Here, $\Gamma_\|$ is also independent of $\delta$, while $\Gamma_\perp$ decays as $1/\delta$.
Interestingly, Eq.~(\ref{draglargedelta}) coincides with the result by Evans and Sackmann for
the drag coefficient of a rigid disk in a passive supported membrane~\cite{evans1988}.

In Fig.~\ref{fig:lift_delta}, we plot $\Gamma_\|$ and $\Gamma_\perp$ in Eqs.~(\ref{eq:drag}) and (\ref{eq:lift}),
respectively, as a function of the odd viscosity difference $\delta$ for various values of $\epsilon$.
As can be seen in Fig.~\ref{fig:lift_delta}(a), $\Gamma_\|$ is almost independent of $\delta$.
However, Fig.~\ref{fig:lift_delta}(b) shows that $\Gamma_\perp$ changes nonmonotonically,
in accordance with Eqs.~(\ref{liftsmalldekta}) and (\ref{liftlargedelta}).
The maximum of $\Gamma_\perp$ shifts to higher values of $\delta$ as $\epsilon$ is increased.

%%%%%%%%%%%%%%%%%%%%
\section{Summary and discussion}
%%%%%%%%%%%%%%%%%%%%
\label{sec:discussion}

In this paper, we have investigated the hydrodynamic forces acting on a 2D liquid domain
that moves laterally in a supported membrane characterized by an odd viscosity.
We combined the momentum decay mechanism of a 2D fluid~\cite{ramachandran2010,seki93,komura95,seki07,sanoop10,sanoop11,sanooppolymer11}
with the concept of odd viscosity~\cite{avron1998}.
Since active rotor proteins can accumulate inside the lipid domain, we have focused on the difference in odd viscosity between the inside and outside of the domain.
Taking into account the momentum decay mechanism of the incompressible 2D fluid,
we have analytically obtained the fluid flow induced by a lateral domain motion.
In the presence of odd viscosity difference, the flow field due to the domain motion is rotated with respect to its direction, as shown in Fig.~\ref{fig:v_x_y}.

Using the obtained flow field, we have calculated the hydrodynamic forces acting on the moving domain.
The resulting drag and lift coefficients are given in Eqs.~(\ref{eq:drag}) and (\ref{eq:lift}).
In contrast to the passive case that does not have an odd viscosity~\cite{saffman1975,saffman1976,evans1988,ramachandran2010},
the existence of a lateral lift force is predicted when the odd viscosity difference is present.
We have discussed in detail the dependence of the drag coefficient $\Gamma_\|$ and lift coefficient $\Gamma_\perp$ on the domain size $\epsilon$ and the odd viscosity difference $\delta$.
The appearance of a finite lift force indicates not only the existence of the odd viscosity, but also its asymmetry between the inside and outside of the domain.

In addition to the asymmetry condition, $\eta_{\rm o}\neq\eta_{\rm o}^\prime$, discussed in this work, we briefly summarize other conditions for finite lift force in incompressible 2D fluids with odd viscosity.
For a laterally moving rigid disk with a nonslip boundary, no lift was observed~\cite{ganeshan2017}, while it was reported to exist within the Oseen approximation~\cite{kogan2016}.
For a bubble with a no-stress boundary condition, lift and torque forces, respectively, emerge for a moving and expanding bubble~\cite{lapa2014,ganeshan2017,souslov2020}.
The forces of rigid disks and bubbles are discussed in more detail below.

Since the governing hydrodynamic equations~(\ref{eq:incomp}) and (\ref{eq:stokes}) are linear in $\mathbf{v}$, the force
$\mathbf{F}$ acting on a circular domain can be generally written as
\begin{align}
\mathbf{F} = -\mathbf{\Gamma}\cdot\mathbf{U},
\label{eq:F}
\end{align}
where $\mathbf{\Gamma}$ is the domain friction tensor and $\mathbf{U}$ is the domain velocity in an arbitrary direction.
Following a similar calculation as before, we find that $\mathbf{\Gamma}$ can be expressed as
\begin{align}
\Gamma_{ij}=
\Gamma_\|\delta_{ij}-\Gamma_\perp\epsilon_{ij},
\label{eq:gamma}
\end{align}
where the coefficients $\Gamma_\|$ and $\Gamma_\perp$ are, respectively, given by Eqs.~(\ref{eq:drag}) and (\ref{eq:lift}) for the simple case $(\eta=\eta^\prime$ and $\lambda=\lambda^\prime)$, or Eqs.~(\ref{eq:generaldrag}) and (\ref{eq:generallift}) for the general case $(\eta\neq\eta^\prime$ and $\lambda\neq\lambda^\prime)$.
When $\delta=0$, the lift coefficient $\Gamma_\perp$ vanishes and the friction tensor satisfies
the reciprocal relation $\Gamma_{ij}=\Gamma_{ji}$.
According to the Lorentz reciprocal theorem~\cite{pozrikidis1992,happel2012,masoud2019},
such a reciprocal property is guaranteed for an arbitrarily shaped object in a passive fluid.
When $\delta\neq0$, the hydrodynamic response becomes nonreciprocal, i.e., $\Gamma_{ij}\neq\Gamma_{ji}$,
leading to a dissipationless lift force.
This is one of the distinctive features of an active chiral fluid characterized by odd viscosity~\cite{hosaka2021}.

In Ref.~\cite{ganeshan2017}, it was shown that a lift force does not exist for an object in an incompressible 2D fluid with odd viscosity when nonslip boundary conditions are imposed.
This is the case when the boundary conditions include only the continuity of velocity as we have
used in Eqs.~(\ref{bcvr})-(\ref{bcvtheta}).
However, in the case of a liquid domain, we also have employed the boundary condition
for the stress continuity as in Eq.~(\ref{bcsigmart}).
Then, the obtained lift force depends on the odd viscosity difference $\delta$.

Some numerical estimates of the physical quantities in the model can be given~\cite{ramachandran2010}.
For a fluid membrane supported by a rigid substrate, the friction parameter in
Eq.~(\ref{eq:balance}) can be identified as $\lambda=\eta_{\rm w}/h$, where
$\eta_{\rm w}$ is the 3D viscosity of the surrounding water and $h$ is the thickness of a thin layer of lubricating water between the membrane and the substrate~\cite{evans1988}.
Then, the hydrodynamic screening length is given by $\kappa^{-1}=(\eta h/\eta_{\rm w})^{1/2}$.
For typical values such as $h\approx10^{-8}$\,m, $\eta_{\rm w}\approx10^{-3}$\,Pa$\cdot$s, and
$\eta\approx10^{-9}$\,Pa$\cdot$s$\cdot$m, we find $\kappa^{-1}\approx10^{-7}$\,m.
Since the size of a lipid domain (raft) is roughly $10$\,nm--$100$\,nm~\cite{simons1997,pike2009}, the
dimensionless domain size $\epsilon=\kappa R$ is estimated to be $0.1 \le \epsilon \le 1$.
Hence, the limiting expressions derived in Eqs.~(\ref{eq:drags}) and (\ref{eq:lifts}) for $\epsilon\ll1$ are the appropriate ones for the drag and lift coefficients.

Next we discuss the value of the domain odd viscosity $\eta_{\rm o}^\prime$ for typical physiological conditions.
Consider the situation where disk-like active rotor proteins concentrate only inside the domain, i.e., $\eta_{\rm o}=0$ and $\eta_{\rm o}^\prime\neq0$, while $\eta=\eta^\prime$ as was assumed above.
In microscopic approaches~\cite{banerjee2017,markovich2021}, it was shown that odd viscosity is related to the angular-momentum density of rotor proteins through the relation, $\eta_{\rm o}^\prime\simeq IT/\zeta$.
Here, $I$ and $T$ are the moment-of-inertia and torque densities, respectively, and $\zeta$ is the rotational friction coefficient of a rotor.

For an active rotor protein of radius $a$ and mass $m$ driven by the torque $\tau$, one can estimate~\cite{evans1988,yang2021} $I=m\rho/\pi$, $T=\rho\tau/(\pi a^2)$, and $\zeta=\eta^\prime\rho/\pi$, which lead to $\eta_{\rm o}^\prime\simeq m\rho\tau/(\pi\eta^\prime a^2)$.
Here, $\rho=N \pi a^2/(\pi R^2)$ is the area fraction of $N$ rotors inside the domain.
Using typical values such as $m\approx10^{-21}$\,kg, $\rho\approx0.3$, $\tau\approx10^{-19}$\,N$\cdot$m, and $a\approx10^{-8}$\,m~\cite{albertsbook,hosaka2017,lenz2004,oppenheimer2019} and assuming that the domain is filled with water ($\eta^\prime\approx10^{-12}$\,Pa$\cdot$s$\cdot$m), we obtain $\eta_{\rm o}^\prime\approx10^{-13}$\,Pa$\cdot$s$\cdot$m.
Then, the odd viscosity ratio is given by $\delta=-\eta_{\rm o}^\prime/\eta\approx-0.1$ and the limiting expressions of Eqs.~(\ref{dragsmalldelta}) and (\ref{liftsmalldekta}) for $|\delta|\ll1$ can be used here for the drag and lift coefficients.

As a special case of a liquid domain, we discuss the hydrodynamic forces acting
on a circular bubble of radius $R$ that moves laterally in an incompressible 2D fluid with odd viscosity.
In Appendix~\ref{sec:bubble}, we obtain the drag and lift coefficients by requiring that $\eta^\prime=0$ and $\eta_{\rm o}^\prime=0$, while $\eta$ and $\eta_{\rm o}$ for the outside of the domain are kept finite.
For a moving bubble, $\Gamma_\|$ and $\Gamma_\perp$ depend on the viscosity ratio $\mu=\eta_{\rm o}/\eta$.
The asymptotic behaviors of the drag and lift coefficients are similar to those of a liquid domain.
In the previous studies, it was reported that the effect of odd viscosity can
be seen as a torque acting on an expanding bubble~\cite{lapa2014,ganeshan2017,souslov2020}.
Our results show that the forces due to odd viscosity exist even for an undeformable object.

In the opposite limit $\eta^\prime\to\infty$, the general drag and lift forces in Eqs.~(\ref{eq:generaldrag}) and (\ref{eq:generallift}) reduce to those acting on a rigid disk.
In this case, the drag coefficient becomes identical to that for a passive supported membrane~\cite{evans1988} as in Eq.~(\ref{draglargedelta}), while the lift coefficient vanishes.
This is reasonable because the boundary conditions at the disk perimeter can be constructed without the stress continuity of Eq.~(\ref{bcsigmart})~\cite{ramachandran2010} and the odd viscosity does not enter in the forces on the disk~\cite{ganeshan2017,souslov2020}.

When the odd viscosity is spatially uniform $(\delta=0)$, it does not affect either the velocity field or the
forces acting on the domain.
This implies that the effect of odd viscosity can be seen in biomembranes when active rotor proteins concentrate locally inside specific domains and the odd viscosity becomes nonuniform.
It would be interesting to investigate experimentally the diffusion of such active
domains by using microrheology techniques~\cite{furst2017}.
When a membrane is in thermal equilibrium, the drag coefficient can be connected to the diffusion
coefficient of the liquid domain through Einstein's relation.
In active fluids, however, such a relation no longer holds and one needs to generalize the
fluctuation-dissipation theorem in the presence of active protein molecules~\cite{hosaka2017,yasuda2017,yasuda2017_2,hosaka2020,hosaka2020_2,hargus2021,yasuda2021}.
Through molecular-dynamics simulations of a particle diffusing in an active chiral fluid, the applicability of Einstein's relation was evaluated~\cite{hargus2021}.
For the Langevin equation with odd viscosity, the asymmetric diffusion tensor is obtained, and is characterized by the ratio of the drag to lift coefficients~\cite{yasuda2021}.
A more detailed discussion of such diffusion phenomena in the active chiral fluid will be given elsewhere~\cite{yasuda2021}.

\begin{acknowledgments}
We thank J.\ E.\ Avron, H.\ Diamant, M.\ Doi, M.\ S.\ Turner, and K.\ Yasuda for useful discussions.
Y.H.\ acknowledges support by a Grant-in-Aid for JSPS Fellows (Grant No.\ 19J20271) from the JSPS.
Y.H.\ also thanks the hospitality of Tel Aviv University, where part of this research was conducted with joint support from Tokyo Metropolitan University and Tel Aviv University.
S.K.\ acknowledges support by a Grant-in-Aid for Scientific Research (C) (Grant No.\ 19K03765) from
the JSPS, and support by a Grant-in-Aid for Scientific Research on Innovative Areas ``Information Physics
of Living Matters'' (Grant No.\ 20H05538) from the Ministry of Education, Culture, Sports, Science and Technology of Japan.
D.A.\ acknowledges support from the Israel Science Foundation (ISF) under Grant No.\ 213/19.
\end{acknowledgments}

\begin{widetext}
\appendix
%%%%%%%%%%%%%%%%%%%%%%%%%%%%%%%
\section{Derivation of the general drag and lift forces}
%%%%%%%%%%%%%%%%%%%%%%%%%%%%%%%
\label{app:c}

The coefficients $C_1, \cdots, C_4, C_1^\prime, \cdots, C_4^\prime$ are determined by the boundary
conditions in Eqs.~(\ref{bcvr})-(\ref{bcsigmart}) and given by
\begin{align}
\begin{aligned}
C_1 &=-RU
\left(\kappa^\prime R I_0 - 2I_1 \right) 
\left[ \eta\kappa^2R^2K_1 + 2(\eta-\eta^\prime)
\left(\kappa R K_0 + 2K_1\right) \right] D_1/(\kappa D)\\
&- \eta^\prime \kappa^{\prime2} R^3 U
\left(\kappa R K_0 + 2K_1\right)I_1 D_1/(\kappa D)
-R^2UK_2D_2/D , \\
C_2 &=2U  \left[ 2(\eta -\eta^\prime) ( \kappa^\prime R I_0 - 2I_1)
+\eta^\prime\kappa^{\prime2} R^2 I_1
 \right]D_1/(\kappa D) + 2U D_2 /(\kappa D),\\
C_3 &= -4\eta (\eta_{\rm o}-\eta_{\rm o}^\prime) \kappa^{\prime2} R^4 U K_1^2 I_2^2/D,\\
C_4 &= -4\eta (\eta_{\rm o}-\eta_{\rm o}^\prime) \kappa^{\prime2} R^3 U  K_1 I_2^2/D,
\end{aligned}
\label{eq:c1234}
\end{align}
and
\begin{align}
\begin{aligned}
C_1^\prime &=U\left[K_0D_2+
\eta ^2 \kappa ^2 {\kappa^\prime}^2 R^4 K_1^2 I_0 I_2+2 \eta  \kappa  R K_0K_1(\kappa^\prime R I_0-I_1)
\left\{ 2(\eta -\eta^\prime)( \kappa^\prime RI_0 - 2I_1) 
 + \eta^\prime \kappa^{\prime2} R^2I_1 \right\} \right.\\
&\left.+ \left\{ 2(\eta -\eta^\prime)( \kappa^\prime RI_0 - 2I_1) 
 + \eta^\prime \kappa^{\prime2} R^2I_1 \right\}^2K_0^2
\right]/D,\\
C_2^\prime &=2 \eta  \kappa  R^2 U K_1 D_1/D,\\
C_3^\prime &= -4 \eta (\eta_{\rm o}-\eta_{\rm o}^\prime) \kappa  {\kappa^\prime} R^2 U  K_0 K_1 I_1 I_2/D,\\
C_4^\prime &= 4 \eta (\eta_{\rm o}-\eta_{\rm o}^\prime) \kappa  {\kappa^\prime} R^3 U  K_0 K_1 I_2/D,
\end{aligned}
\label{eq:c1234p}
\end{align}
where
\begin{align}
\begin{aligned}
D &= D_1^2+K_0D_2, \\
D_1 &=
2(\eta -\eta^\prime) ({\kappa^\prime} R   I_0 - 2I_1) K_0
+ {\kappa^\prime} R^2
 \left(\eta^\prime {\kappa^\prime} K_0 I_1 + \eta\kappa K_1 I_2 \right), \\
D_2 &= 4({\eta_{\rm o}}-{\eta_{\rm o}^\prime})^2 {\kappa^\prime}^2 R^2  K_0 I_2^2.
\end{aligned}
\label{eq:D12}
\end{align}
In the above, we have used the notations
$K_n = K_n(\kappa R) = K_n(\epsilon)$ and $I_n = I_n(\kappa^\prime R) = I_n(\epsilon^\prime)$.
In the main text, we consider the case $\eta=\eta^\prime$ and $\lambda=\lambda^\prime$ (or equivalently $\kappa=\kappa^\prime$), and the function $I_n(\kappa R)=I_n(\epsilon)$ is written as $I_n$.

Substituting $C_1$ and $C_2$ into Eq.~(\ref{eq:Fx}) and $C_3$ and $C_4$ into Eq.~(\ref{eq:Fy}), we obtain
the general drag and lift forces as
\begin{align}
\frac{F_x}{4\pi\eta}&=
U \epsilon (\delta {\epsilon^\prime}I_2)^2 (\epsilon  K_0+4 K_1)K_0/M
+ (U \epsilon/4)
\left[ 
2(\nu-1)K_0 (2I_1-\epsilon^\prime I_0)
+ \epsilon^\prime \left( \epsilon K_1I_2 + \nu\epsilon^\prime K_0I_1 \right)
 \right] \nonumber\\
&\times
\left[ \left\{ \nu{\epsilon^\prime}^2I_1 + 2(\nu-1)(2I_1-\epsilon^\prime I_0) \right\} (\epsilon K_0 + 4K_1) - \epsilon^2 K_1(2I_1-\epsilon^\prime I_0) \right]/M ,\label{eq:generaldrag}
\end{align}
and
\begin{align}
\frac{F_y}{4\pi\eta}&=
2U \delta  (\epsilon \epsilon^{\prime} K_1 I_2 )^2/M,
\label{eq:generallift}
\end{align}
where $\nu=\eta^\prime/\eta$, $\delta=(\eta_{\rm o}-\eta_{\rm o}^\prime)/\eta$, and
\begin{align}
M &= \left[ 2\left(\nu-1 \right) K_0 \left( 2I_1 - \epsilon^\prime I_0 \right)  + \epsilon^\prime \left(\epsilon K_1I_2 + \nu\epsilon^\prime K_0I_1 \right) \right]^2
+4 (\delta{\epsilon^\prime} K_0 I_2)^2.
\end{align}
When $\eta=\eta^\prime$ and $\lambda=\lambda^\prime$ (or equivalently $\nu=1$ and $\epsilon=\epsilon^\prime$),
we obtain Eqs.~(\ref{eq:drag}) and (\ref{eq:lift}).

%%%%%%%%%%%%%%%%%%%%%%%%%%%%%%%%%%%%%%%%%%
\section{Drag coefficient for a 2D liquid domain when $\eta_{\rm o}=\eta_{\rm o}^\prime=0$}
%%%%%%%%%%%%%%%%%%%%%%%%%%%%%%%%%%%%%%%%%%
\label{app:sanoop}

We summarize the passive case without odd viscosity, which was studied in Ref.~\cite{ramachandran2010}.
When $\eta_{\rm o}=\eta_{\rm o}^\prime=0$ (while $\nu\neq1$ or $\eta\neq\eta^\prime$), the coefficients, $C_3,C_4,C_3^\prime,$ and $C_4^\prime,$ become zero, as can be seen in Eqs.~(\ref{eq:c1234})-(\ref{eq:D12}).
Then, the scalar and vector potentials in Eqs.~(\ref{eq:phi}), (\ref{eq:A}), (\ref{eq:phip}), and (\ref{eq:Ap}) reduce to
\begin{align}
\phi =\frac{C_{1}}{r}\cos\theta,~~~~
A =C_{2} K_{1}(\kappa r) \sin \theta,~~~~
\phi^{\prime} =C_{1}^{\prime} r \cos \theta,~~~~
A^{\prime}=C_{2}^{\prime} I_{1}\left(\kappa^{\prime} r\right) \sin \theta,
\label{eq:potentials0}
\end{align}
respectively.
Calculating the corresponding velocity fields and stress tensors and applying the boundary conditions of Eqs.~(\ref{bcvr})-(\ref{bcsigmart}), one can obtain the drag and lift coefficients as
\begin{align}
\frac{\Gamma_\|}{4 \pi \eta}=\frac{\epsilon^{2}}{4}
+\frac{\epsilon K_{1}\left[\nu\left(4+\epsilon^{\prime 2}\right) I_{1}-2 \nu\epsilon^{\prime} I_{0}+2 \left(\epsilon^{\prime} I_{0}-2 I_{1}\right)\right]}
{\nu K_{0}\left[\left(4+\epsilon^{\prime 2}\right) I_{1}-2 \epsilon^{\prime} I_{0}\right]
+\left(2 K_{0}+\epsilon K_{1}\right)\left(\epsilon^{\prime} I_{0}-2 I_{1}\right)},
\label{eq:drag0}
\end{align}
and $\Gamma_\perp=0$, respectively.
When $\nu=1$ or $\eta=\eta^\prime$, Eq.~(\ref{eq:drag0}) coincides with the drag coefficient derived in Eq.~(\ref{dragsmalldelta}) for $|\delta|\ll1$.

%%%%%%%%%%%%%%%%%%%%%%%%%%%%%
\section{Drag and lift coefficients for a 2D bubble}
%%%%%%%%%%%%%%%%%%%%%%%%%%%%%
\label{sec:bubble}

We derive the hydrodynamic forces acting on a moving bubble of radius $R$.
By setting $\eta^\prime=0$ and $\eta_{\rm o}^\prime=0$ in
Eqs.~(\ref{eq:generaldrag}) and (\ref{eq:generallift}), we obtain the drag and lift coefficients as
\begin{align}
\begin{aligned}
\frac{\Gamma_\|}{4\pi\eta} &=
\frac{\epsilon^2}{4}+\frac{2\epsilon K_1}{2K_0+\epsilon K_1}
\left[1+\frac{2\mu^2\epsilon K_0K_1}{(2K_0+\epsilon K_1)^2+4(\mu K_0)^2} \right],\\
\frac{\Gamma_\perp}{4\pi\eta} &=
\frac{2 \mu (\epsilon K_1)^2}
{(2K_0+\epsilon K_1)^2+4(\mu K_0)^2},
\end{aligned}
\end{align}
with $\mu=\eta_{\rm o}/\eta$.
In the limits of $\epsilon\ll1$ and $\epsilon\gg1$, we obtain respectively for arbitrary $\mu$
\begin{align}
\begin{aligned}
\frac{\Gamma_\|}{4\pi\eta} &\approx
\frac{\ln(2/\epsilon)-\gamma+1/2+\mu^2[\ln(2/\epsilon)-\gamma]}
{[\ln(2/\epsilon)-\gamma+1/2]^2+\mu^2[\ln(2/\epsilon)-\gamma]^2},\\
\frac{\Gamma_\perp}{4\pi\eta} &\approx
\frac{\mu}{2[\ln(2/\epsilon)-\gamma+1/2]^2+2\mu^2[\ln(2/\epsilon)-\gamma]^2},
\end{aligned}
\end{align}
and
\begin{align}
\begin{aligned}
\frac{\Gamma_\|}{4\pi\eta} &\approx \frac{\epsilon^2}{4},\\
\frac{\Gamma_\perp}{4\pi\eta} &\approx 2\mu.
\end{aligned}
\end{align}

\end{widetext}

%\bibliography{myref}

%\begin{comment}
%apsrev4-2.bst 2019-01-14 (MD) hand-edited version of apsrev4-1.bst
%Control: key (0)
%Control: author (8) initials jnrlst
%Control: editor formatted (1) identically to author
%Control: production of article title (0) allowed
%Control: page (0) single
%Control: year (1) truncated
%Control: production of eprint (0) enabled
%
%\end{comment}

\end{document}